\documentclass[11pt]{article}
\usepackage{geometry}                
\usepackage{graphicx}
\usepackage{amssymb}
\usepackage{epstopdf}
\usepackage{flushend}
\usepackage{url}
\usepackage{multirow}
\usepackage[flushleft]{threeparttable}

\newcommand{\Xknndtw}{KNN-DTW}
\newcommand{\Xsaxvsm}{SAX-VSM}

\title{Early Detection of Promoted Campaigns on Social Media}
\author{Onur Varol$^1$, Emilio Ferrara$^{2}$\footnote{Corresponding author: emiliofe@usc.edu}, Filippo Menczer$^1$, Alessandro Flammini$^1$}
\date{
	$^1$ School of Informatics and Computing, Indiana University \\
	$^2$ Information Sciences Institute, University of Southern California \\
}

\begin{document}
\maketitle
\begin{abstract} 
Social media expose millions of users every day to information campaigns --- some emerging organically from grassroots activity, others sustained by advertising or other coordinated efforts. These campaigns contribute to the shaping of collective opinions. While most information campaigns are benign, some may be deployed for nefarious purposes, including terrorist propaganda, political astroturf, and financial market manipulation. It is therefore important to be able to detect whether a meme is being artificially promoted at the very moment it becomes wildly popular. This problem has important social implications and poses numerous technical challenges. As a first step, here we focus on discriminating between trending memes that are either organic or promoted by means of advertisement. The classification is not trivial: ads cause bursts of attention that can be easily mistaken for those of organic trends. We designed a machine learning framework to classify memes that have been labeled as trending on Twitter.After trending, we can rely on a large volume of activity data. Early detection, occurring immediately at trending time, is a more challenging problem due to the minimal volume of activity data that is available prior to trending.Our supervised learning framework exploits hundreds of time-varying features to capture changing network and diffusion patterns, content and sentiment information, timing signals, and user meta-data. We explore different methods for encoding feature time series. Using millions of tweets containing trending hashtags, we achieve 75\%  AUC score for early detection, increasing to above 95\% after trending. We evaluate the robustness of the algorithms by introducing random temporal shifts on the trend time series. Feature selection analysis reveals that content cues provide consistently useful signals; user features are more informative for early detection, while network and timing features are more helpful once more data is available.
\end{abstract}

\section{Introduction}
An increasing number of people rely, at least in part, on information shared on social media to form opinions and make choices on issues related to lifestyle, politics, health, and products purchases~\cite{bakshy2011everyone,bond201261,olteanu2017distilling}. Such reliance provides a variety of entities --- from single users to corporations, interest groups, and governments --- with motivation to influence collective opinions through active participation in online conversations. There are also obvious incentives for the adoption of covert methods that enhance both perceived and actual popularity of promoted information. 
There are abundant recently reported examples of abuse: astroturf in political campaigns, or attempts to spread fake news through social bots under the pretense of grassroots conversations~\cite{ratkiewicz2011detecting,ferrara2016rise,bessi2016social}; pervasive spreading of unsubstantiated rumors and conspiracy theories~\cite{bessi2015science};  orchestrated boosting of perceived consensus on relevant social issues performed by governments~\cite{guardian2015}; 
propaganda and recruitment by terrorist organizations, like ISIS~\cite{berger2015isis,ferrara2016predicting}; 
and actions involving social media and stock market manipulation~\cite{SEC2015}.

The situation is ripe with dangers as people are rarely equipped to recognize propaganda or promotional campaigns as such. It can be difficult to establish the origin of a piece of news, the reputation of its source, and the entity behind its promotion on social media, due both to the intrinsic mechanisms of sharing and to the high volume of information that competes for our attention. Even when the intentions of the promoter are benign, we easily interpret large (but possibly artificially enhanced) popularity as widespread endorsement of, or trust in, the promoted information. 

There are at least three questions about information campaigns that present scientific challenges: what, how, and who. The first concerns the subtle notion of trustworthiness of information, ranging from verified facts~\cite{ciampaglia2015computational}, to rumors and exaggerated, biased, unverified or fabricated news~\cite{ratkiewicz2011detecting,zhao2015enquiring,bessi2015science}. The second considers the tools employed for the propaganda. Again, the spectrum is wide: from a known brand that openly promotes its products by targeting users that have shown interest, to the adoption of social bots, trolls and fake or manipulated accounts that pose as humans \cite{varol2017online,ferrara2016rise,davis2016botornot,clark2015vaporous,haustein2016tweets}. The third question relates to the (possibly concealed) entities behind the promotion efforts and the transparency of their goals. 
Even before these question can be explored, one would need to be able to \emph{identify} an information campaign in social media. 
But discriminating such campaigns from grassroots conversations poses both theoretical and practical challenges. Even the very definition of ``campaign'' is conceptually difficult, as it entangles the nature of the content (e.g., product or news), purpose of the source (e.g., deception, recruiting), strategies of dissemination (e.g., promotion or orchestration), different dynamics of user engagement (e.g., the aforementioned social bots), and so on. 

This paper takes a first step toward the development of computational methods for the \emph{early detection} of information campaigns. In particular, we focus on trending memes and on a special case of promotion, namely advertisement, because they provide convenient operational definitions of social media campaigns. We formally define the task of discriminating between organic and promoted trending memes. Future efforts will aim at extending this framework to other types of information campaign.

\subsection{The challenge of identifying promoted content}

\begin{table} [!t]
\centering
\caption{Summary statistics of collected data about promoted and organic trends on Twitter.}
\begin{tabular}{lcccc}
\hline
    &  \multicolumn{2}{c}{Promoted} &  \multicolumn{2}{c}{Organic} \\ 
\hline
Dates & \multicolumn{2}{c}{1 Jan-- 31 Apr 2013} & \multicolumn{2}{c}{1--15 Mar 2013}\\
No. trends 					& \multicolumn{2}{c}{75} & \multicolumn{2}{c}{852} \\
& mean & st. dev. & mean & st. dev.\\
\cline{2-5}
Avg. no. tweets      & 2,385 & 6,138 & 3,692 & 9,720 \\
Avg. no. unique users   & 2,090 & 5,050 & 2,828 & 8,240 \\
Avg. retweet ratio      & 42\% & 13.8\% & 33\% & 18.6\% \\
Avg. reply ratio            & 7.5\% & 7.8\% & 20\% & 21.8\% \\
Avg. no. urls           & 0.25 & 0.176 & 0.15 & 0.149 \\
Avg. no. hashtags       & 1.7 & 0.33 & 1.7 & 0.78 \\
Avg. no. mentions       & 0.8 & 0.28 & 0.9 & 0.35 \\
Avg. no. words          & 13.5 & 2.21 & 12.2 & 2.74 \\
\hline
\end{tabular}
\label{tab:dataset}
\end{table}

On Twitter, it is common to observe \emph{hashtags} --- keywords preceded by the \# sign that identify messages about a specific topic --- enjoying sudden bursts in activity volume due to intense posting by many users with an interest in the topic~\cite{lehmann2012dynamical,yang2011patterns,myers2014bursty}. Such hashtags are labeled as \emph{trending} and are highlighted on the Twitter platform. Twitter algorithmically identifies trending topics in a predetermined set of geographical locations. Although Twitter recently included personalized and clustered trends, the ones in the collection analyzed here correspond to single hashtags selected on the basis of their popularity. Unfortunately, detailed knowledge about the algorithm and criteria used to identify organic trends is not publicly available~\cite{twittertrend}. 
Other hashtags are exposed prominently after the payment of a fee by parties that have an interest in enhancing their popularity. Such hashtags are called \emph{promoted} and often enjoy subsequent bursts of popularity similar to those  of trending hashtags, therefore being listed among trending topics. 

Of course, once Twitter labels a hashtag as trending, it is not necessary to detect whether or not it is promoted --- this information is disclosed by Twitter. However, since it is difficult to manually annotate a sufficiently large datasets of campaigns, we use organic and promoted trending topics as a \emph{proxy} for a broader set of campaigns, where promotion mechanisms may be hidden.  
Our data collection methodology provide us with a large source of reliable ``ground truth'' labels about promotion, which represent an ideal testbed to evaluate detection algorithms. These algorithms have to determine whether or not a hashtag is promoted based on information that would be available even in cases where the nature of a trend is unknown. 
We stress that our goal of distinguishing mechanisms for promoting popular content is different from that of predicting viral topics, an interesting area of research in its own right~\cite{weng2012virality, cheng2014can, cheng2016cascades}.

\begin{figure}[!t] \centering
\includegraphics[width=.75\textwidth]{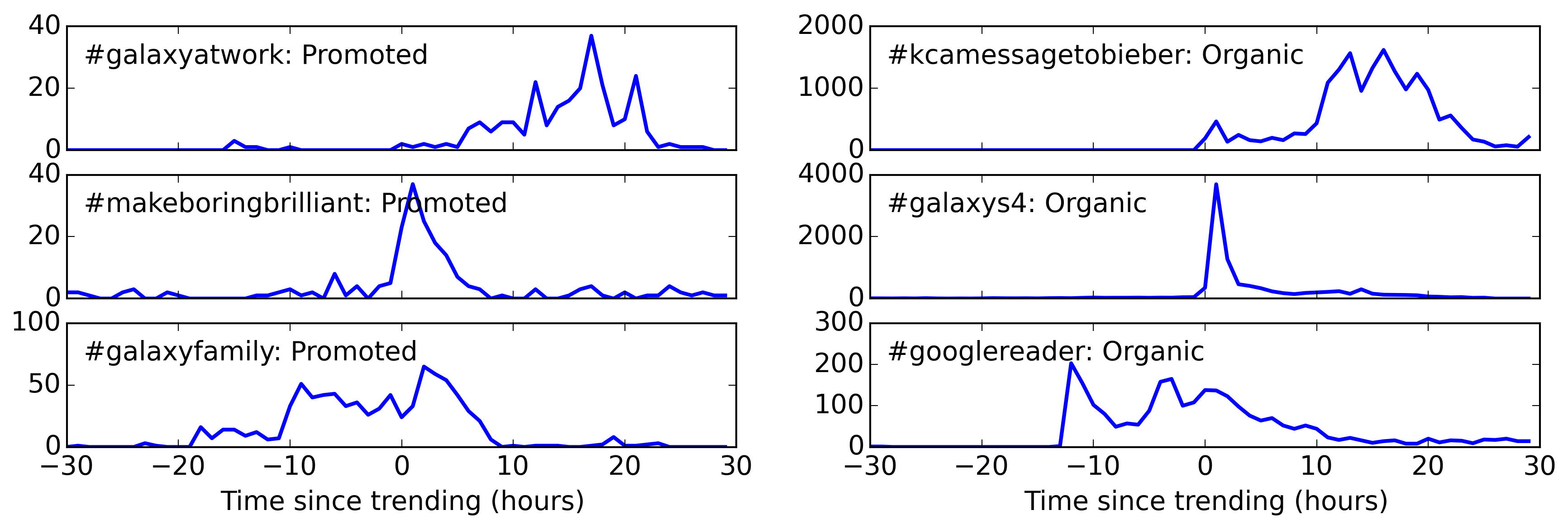}
\caption{Time series of trending hashtags. Comparison of the time series of the volume (number of tweets per hour in our sample) relative to promoted (left) and organic (right) trends with similar temporal dynamics.} 
\label{fig:example}
\end{figure}

Discriminating between promoted and organically trending topics  
is not trivial, as Table~\ref{tab:dataset} illustrates --- promoted and organic trending hashtags often have similar  characteristics. 
One might assume that promoted trends display volume patterns characteristic of exogenous influence, with sudden bursts of activity, whereas organic trends would conform to more gradual volume growth patterns typical of endogenous processes~\cite{sornette2004endogenous,myers2012information,lehmann2012dynamical}. However, Fig.~\ref{fig:example} shows that promoted and organic trends exhibit similar volume patterns over time. 
Furthermore, promoted hashtags may preexist the moment in which they are given the promoted status and may have originated in an entirely grassroots fashion. It is therefore conceivable for such hashtags to display features that are largely indistinguishable from those of other grassroots hashtags about the same topic, at least until the moment of promotion. 

\begin{figure} [!t]
\centering
\includegraphics[width=0.75\columnwidth]{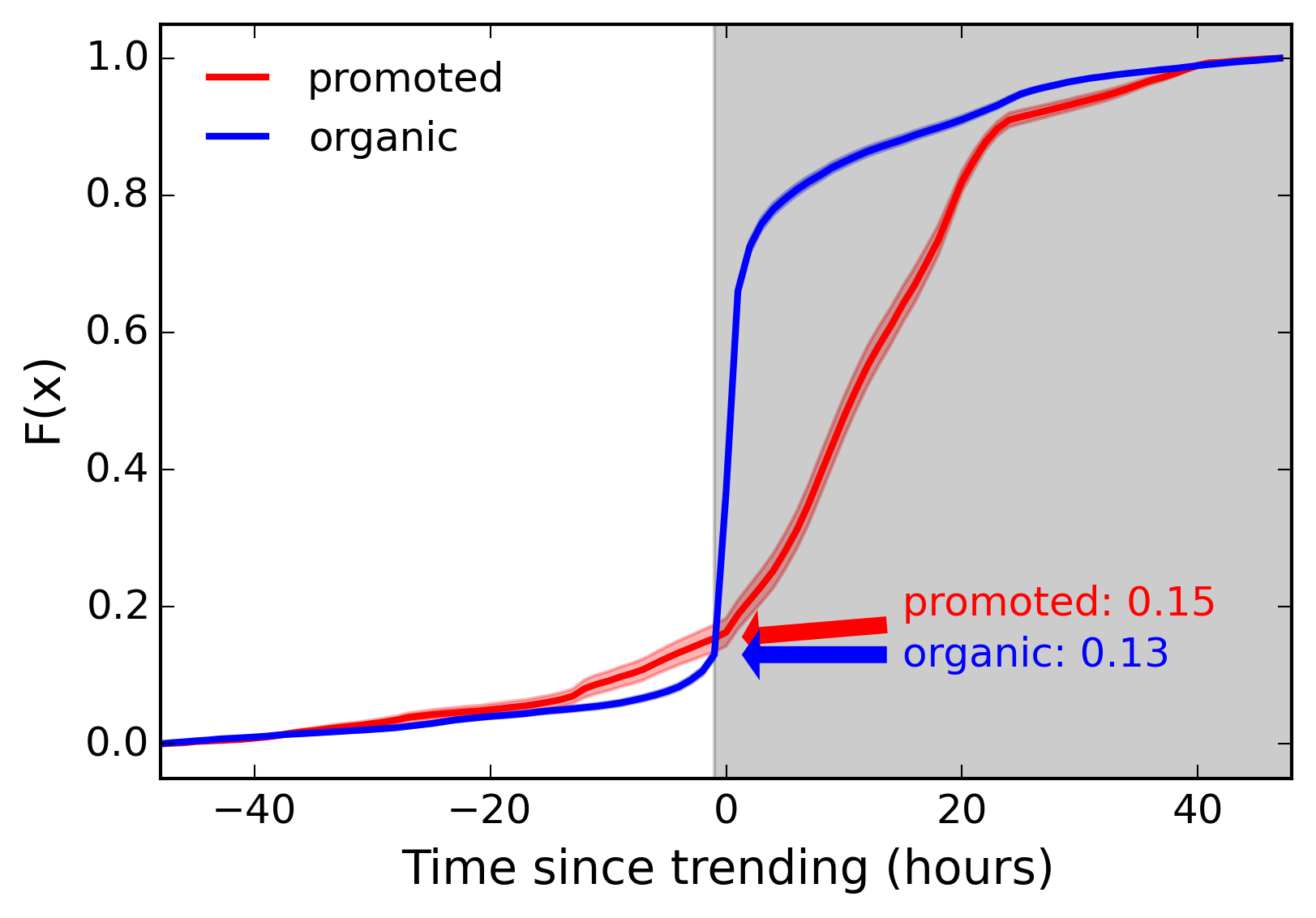}
\caption{Cumulative fraction of tweets as a function of time. On average, only 13\% of the tweets in the organic class and 15\% of the tweets in the promoted class are produced prior to the trending point. The majority of tweets are observed after the trending point, with a rapid increase around trending time. }
\label{fig:cumulative-fraction}
\end{figure}

The analysis in this paper is motivated by the goal of identifying promoted campaigns at the earliest possible time. The early detection task addresses the difficulty of judging the nature of a hashtag using only the limited data available immediately before trending. 
Fig.~\ref{fig:cumulative-fraction} illustrates the shortage of information available for early detection. 
It is also conceivable that once the promotion has triggered interest in a hashtag, the conversation is sustained by the same mechanisms that characterize organic diffusion. Such noise around popular conversations may present an added difficulty for the early detection task.

\subsection{Contributions and outline}

The major contribution of this paper, beyond formulating the problem of detection of campaigns in social media, is the development and validation of a supervised machine learning framework that takes into consideration the temporal sequence of messages associated with a trending hashtag on Twitter and successfully classifies it as either ``promoted'' (advertised) or ``organic'' (grassroots). 
The proposed framework adopts time-varying features built from network structure and diffusion patterns, language, content and sentiment information, timing signals, and user meta-data. 
In the following sections we discuss the data we collected and employed, the procedure for feature extraction and selection, the implementation of the learning framework, and the evaluation of our system.

\section{Data and methods}

\begin{figure}[t!]
\centering
\includegraphics[width=.2\columnwidth]{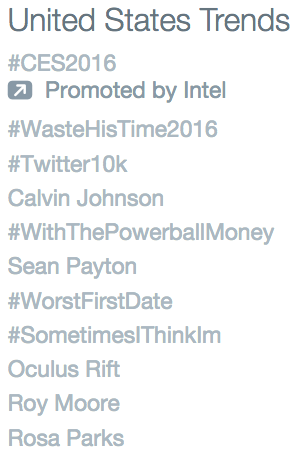}
\caption{Screenshot of Twitter U.S. trends taken on Jan. 6, 2016. The hashtag \texttt{\#CES2016} was promoted on this date.}
\label{fig:usa_trends}
\end{figure}

\subsection{Dataset description} 
\label{sub:dataset}

The dataset adopted in this study consists of Twitter posts (\emph{tweets}) that contain a trending hashtag and appeared during a defined observation period.
Twitter provides an interface that lists trending topics, with clearly labeled \textit{promoted} trends at the top (Fig.~\ref{fig:usa_trends}). We crawled the Twitter webpage at regular intervals of 10 minutes to collect all organic and promoted hashtags trending in the United States between January and April 2013, for a total of $N=927$ hashtags. This constitutes our ground-truth dataset of \emph{promoted} and \emph{organic} trends.

We extracted a sample of organic trends observed during the first two weeks of March 2013 for our analysis. While Twitter allows for at most one promoted hashtag per day, dozens of organic trends appear in the same period. 
As a result, our dataset is highly imbalanced, with the promoted class more than ten times smaller than the the organic one (cf. Table~\ref{tab:dataset}). 
Such an imbalance, however, reflects our expectation to observe in the Twitter stream a minority of promoted conversations blended in a majority of organic content. Therefore we did not balance the classes by resampling, to study the campaign detection problem under realistic conditions. 

Hashtags may trend multiple times on Twitter. However, those in our collection only trended once during our observation period. For each trend, we retrieved all tweets containing the trending hashtag from an archive containing a 10\% random sample of the public Twitter stream. 
The collection period was hashtag-specific: for each hashtag we obtained all tweets produced in a four-day interval, starting two days before its trending point and extending to two days after that. 
This procedure provides an extensive coverage of the temporal history of each trending hashtag in our dataset and its related tweets, 
allowing us to study the characteristics of each trend before, during, and after the trending point.

Given that each trend is described by a collection of tweets over time, we can aggregate data in sliding time windows $[t, t+\ell)$ of duration $\ell$ and compute features on the subsets of tweets produced in these windows. 
A window can slide by time intervals of duration $\delta$. The next window 
therefore contains tweets produced in the interval $[t+\delta, t+\ell+\delta)$.  
We experimented with various time window lengths and sliding parameters, and the optimal performance is often obtained with  windows of duration $\ell=6$ hours sliding by $\delta=20$ minutes. 

We have made the IDs of all tweets involved in the trending hashtags analyzed in this paper available in a public dataset.\footnote{\url{carl.cs.indiana.edu/data/ovarol/trend-dataset.tar.gz}}

\subsection{Features} 
\label{sub:features}

Our framework computes features from a collection of tweets in some time interval. 
The system generates 487 features in five different classes: network structure and information diffusion patterns, content and language, sentiment, timing, and user meta-data. The classes and types of features are reported in Table~\ref{tab:features} and discussed next. 
All of the feature time series in this study are available in our public dataset.

\begin{table} [!t]
\centering \tiny
\caption{List of 487 features extracted by our framework. } 
\begin{threeparttable}
\begin{tabular}{llc}
\hline
Class & Feature description & No. of features\\
\hline\multirow{9}{*}{\textbf{Network($\dag$)}}
& Number of nodes & 1 \\
& Number of edges & 1 \\
& (*) Strength distribution & 8 \\
& (*) In-strength distribution & 8 \\
& (*) Out-strength distribution & 8\\  
& (*) Distribution of number of nodes in the connected components & 8 \\ 
& Network density of whole and largest connected component & 2 \\
& Network assortativity of whole and largest connected component & 2 \\
& Mean shortest path length of the largest connected component & 1 \\

\hline\multirow{10}{*}{\textbf{User}}
& (*) Sender's follower count & 8 \\
& (*) Sender's followee count & 8 \\
& (*) Sender's number of favorite tweets & 8 \\
& (*) Sender's number of Twitter statuses posted & 8 \\
& (*) Sender's number of lists subscribed to & 8 \\
& (*) Originator's follower count & 8 \\
& (*) Originator's followee count & 8 \\
& (*) Originator's number of favorite tweets & 8 \\
& (*) Originator's number of Twitter statuses posted & 8 \\
& (*) Originator's number of lists subscribed to & 8 \\

\hline\multirow{4}{*}{\textbf{Timing}}
& Number of tweets appeared in a given window & 1 \\
& (*) Time between two consecutive tweets & 8 \\
& (*) Time between two consecutive retweets & 8 \\
& (*) Time between two consecutive mentions & 8 \\

\hline\multirow{7}{*}{\textbf{Content}}
& (*) Number of hashtags in a tweet & 8  \\
& (*) Number of mentions in a tweet & 8  \\
& (*) Number of URLs in a tweet & 8  \\
& (*,**) Frequency of POS tags in a tweet & 64 \\
& (*,**) Proportion of POS tags in a tweet & 64 \\
& (*) Number of words in a tweet & 8 \\
& (*) Entropy of words in a tweet & 8  \\

\hline\multirow{18}{*}{\textbf{Sentiment}}
& (***) Happiness scores of aggregated tweets & 2 \\
& (***) Valence scores of aggregated tweets & 2  \\
& (***) Arousal scores of aggregated tweets & 2 \\
& (***) Dominance scores of single tweets & 2 \\
& (*) Happiness score of single tweets & 8 \\
& (*) Valence score of single tweets & 8 \\
& (*) Arousal score of single tweets & 8 \\
& (*) Dominance score of single tweets & 8 \\
& (*) Polarization score of single tweets & 8 \\
& (*) Entropy of polarization scores of single tweets & 8 \\
& (*) Positive emoticons entropy of single tweets & 8  \\
& (*) Negative emoticons entropy of single tweets & 8  \\
& (*) Emoticons entropy of single tweets & 8  \\
& (*) Ratio between positive and negative score of single tweets & 8  \\
& (*) Number of positive emoticons in single tweets & 8  \\
& (*) Number of negative emoticons in single tweets  & 8 \\
& (*) Total number of emoticons in single tweets & 8  \\
& Ratio of tweets that contain emoticons & 1\\
\hline
\end{tabular} 
\begin{tablenotes}
\item[$\dag$] We consider three types of network: retweet, mention, and hashtag co-occurrence networks. The hashtag co-occurrence network is undirected.
\item[*] Distribution types. For each distribution, the following eight statistics are computed and used as individual features: min, max, median, mean, std. deviation, skewness, kurtosis, and entropy. 
\item[**] Part-of-Speech (POS) tag. There are eight POS tags: verbs, nuns, adjectives, modal auxiliaries, pre-determiners, interjections, adverbs, and pronouns.
\item[***] For each feature we compute mean and std. deviation. 
\end{tablenotes}
\end{threeparttable}
\label{tab:features}
\end{table}

\subsubsection{Network and diffusion features}

Twitter actively fosters interconnectivity. Users are linked by means of \emph{follower/followee} relations. Content travels from person to person via \emph{retweets}. Tweets themselves can be addressed to specific users via \emph{mentions}. The network structure carries crucial information for the characterization of different types of communication. In fact, the usage of network features significantly helps in tasks like astroturf detection \cite{ratkiewicz2011detecting}. Our system reconstructs three types of networks: retweet, mention, and hashtag co-occurrence networks. Retweet and mention networks have users as nodes, with a directed link between a pair of users that follows the direction of information spreading --- toward the user retweeting or being mentioned.
Hashtag co-occurrence networks have undirected links between hashtag nodes when two hashtags have occurred together in a   tweet. 
All networks are weighted according to the number of interactions and co-occurrences. 
For each network, a set of features is computed, including in- and out-strength (weighted degree) distribution, density, shortest-path distribution, and so on. (cf. Table~\ref{tab:features}).

\subsubsection{User-based features}

User meta-data is crucial to classify communication patterns in social media \cite{mislove2011understanding,ferrara2016rise}.
We extract user-based features from the details provided by the Twitter API about the author of each tweet and the originator of each retweet. Such features include the distribution of follower and followee numbers, and the number of tweets produced by the users (cf. Table~\ref{tab:features}).

\subsubsection{Timing features}

The temporal dimension associated with the production and consumption of content may reveal important information about campaigns and their evolution~\cite{Ghosh11snakdd}. 
The most basic time-related feature we considered is the number of tweets produced in a given time interval. Other timing features describe the distributions of the intervals between two consecutive events, like two tweets or retweets (cf. Table~\ref{tab:features}). 

\subsubsection{Content and language features}

Many recent papers have demonstrated the importance of content and language features in revealing the nature of social media conversations~\cite{danescu2013no,mcauley2013amateurs,mocanu2013twitter,botta2015quantifying,letchford2015advantage}. For example, deceiving messages generally exhibit informal language and short sentences \cite{briscoe2014cues}.
Our system extracts language features by applying a \emph{Part-of-Speech} (POS) tagging technique, which identifies different types of natural language components, or \emph{POS tags}. 
The following POS tags are extracted: verbs, nouns, adjectives, modal auxiliaries, pre-determiners, interjections, adverbs, pronouns, and wh-pronouns.\footnote{\url{www.comp.leeds.ac.uk/ccalas/tagsets/upenn.html}} 
Tweets can be therefore analyzed to study how such POS tags are distributed. 
Other content features include the length and entropy of the tweet content (cf. Table~\ref{tab:features}).

\subsubsection{Sentiment features}

Sentiment analysis is a powerful tool to describe the attitude or mood of an online conversation.
Sentiment extracted from social media conversations has been used to forecast offline events, including elections and financial market fluctuations~\cite{tumasjan2010predicting,bollen2011twitter}, and is known to affect information spreading \cite{mitchell2013geography,ferrara2015quantifying}. 
Our framework leverages several sentiment extraction techniques to generate various sentiment features, including \emph{happiness score}~\cite{kloumann2012positivity}, \emph{arousal, valence and dominance scores}~\cite{warriner2013norms}, \emph{polarization and strength}~\cite{wilson2005recognizing}, and \emph{emotion score}~\cite{agarwal2011sentiment} (cf. Table~\ref{tab:features}).

\subsection{Feature selection} 
\label{sub:selection}

Our system generates a set $I$ of $|I|=487$ features (cf. Table \ref{tab:features}) designed to extract signals from a collection of tweets and distinguish promoted trends from organic ones.
Some features are more predictive than others; some are by definition correlated with each other due to temporal dependencies. Most of the correlations are related to the volume of data. For instance the two most correlated features immediately prior to the trending point are the size of the hashtag cooccurrence network and the size of its largest connected component (Pearson's $\rho=0.75$). This is why it is important to perform feature selection to eliminate redundant features and identify a combination of features that yield good classification performance. 

There are several methods to select the most predictive features in a classification task \cite{guyon2003introduction}. We implemented a simple greedy forward feature selection method, summarized as follows: (i)~initialize the set of selected features $S = \emptyset$; (ii)~for each feature $i \in I-S$, consider the union set $U = S \cup \{i\}$; (iii)~train the classifier using the features in $U$; (iv)~test the average performance of the classifier trained on this set; (v)~add to $S$ the feature that provides the best performance; (vi)~repeat (ii)--(v). 
We terminate the feature selection procedure if the AUC (cf. Sec.~\ref{sub:algorithms}) increases by less than 0.05 between two consecutive steps. Most of the experiments terminate after selecting fewer than 10 features. The time series for the selected features are passed as input to the  learning algorithms. In the next subsections we provide details about our experimental setting and learning models.

\subsection{Experimental setting}
\label{sub:exp_setting}

\begin{figure}[!t]
\centering
\includegraphics[width=\columnwidth]{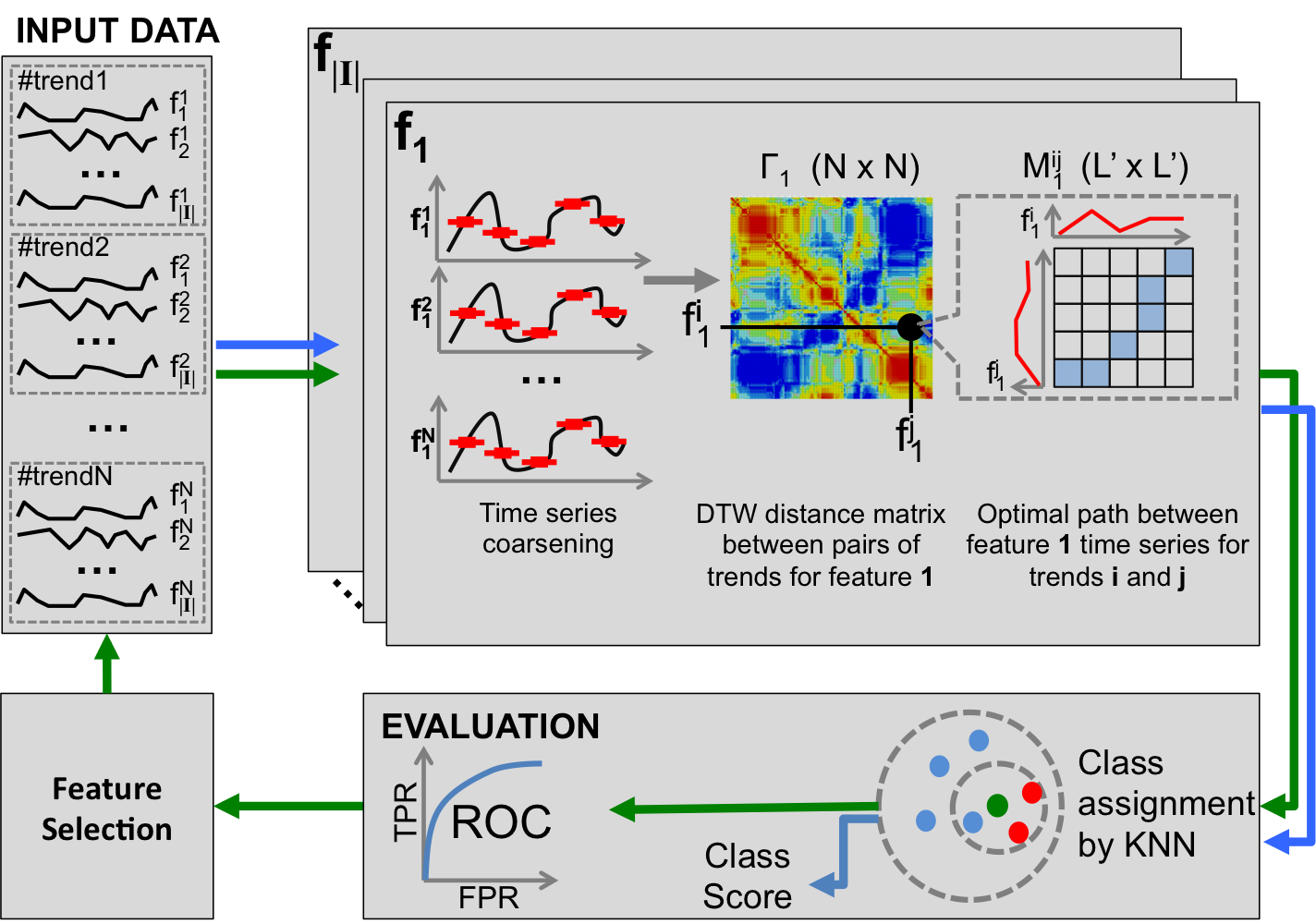}
\caption{Wrapper method description for \Xknndtw. We present the pipeline of our complete system, including feature selection and model evaluation steps. Input data feed into the system for training (green arrow) and testing (blue arrow) steps.}
\label{fig:method-descriptions}
\end{figure}

Our experimental setting follows a pipeline of feature selection, model building, and performance evaluation. We apply the \emph{wrapper} approach to select features and evaluate performance iteratively~\cite{john1994irrelevant}.
During each iteration (Fig.~\ref{fig:method-descriptions}), we train and evaluate models using candidate subsets of features and expand the set of selected features using the greedy approach described in Sec.~\ref{sub:selection}. Once we identify the set of features that performs best, we report results of experiments using only this set of features. 

In each experiment and for each feature, an algorithm receives in input a time series with $L=35$ data points to carry out its detection. The length of the time series and its delay $D$ with respect to the trending point are discussed in Sec.~\ref{sec:results}; different experiments will consider different delays. 

A set of feature time series is used to either train a learning model or evaluate its accuracy. The learning algorithms are discussed in the next subsection. For evaluation, we compute a Receiver Operating Characteristic (ROC) curve, which plots the true positive rate (TPR) versus the false positive rate (FPR) at various thresholds. Accuracy is evaluated by measuring the Area Under the ROC Curve (AUC) \cite{fawcett2006introduction} with 10-fold cross validation, and averaging AUC scores across the folds. A random-guess classifier produces the diagonal line where TPR equals FPR, corresponding to a 50\% AUC score. Classifiers with higher AUC scores perform better and the perfect classifier in this setting achieves a 100\% AUC score. We adopt AUC to measure accuracy because it is not biased by the imbalance in our classes (75 promoted trends versus 852 organic ones, as discussed earlier). 

\subsection{Learning algorithms}
\label{sub:algorithms}

Let us describe the learning systems for online campaign detection based on multidimensional time-series data from social media. 
We identified an algorithm, called \emph{K-Nearest Neighbor with Dynamic Time Warping} ({\Xknndtw}), that is capable of dealing with multidimensional time series classification. For evaluation purposes, we compare the classification results against two baselines: {\Xsaxvsm}
and KNN. These three methods are described next. 

\subsubsection{{\Xknndtw} classifier}
\label{sub:knndtw}

{\Xknndtw} is a state-of-the-art algorithm to classify multidimensional time series, illustrated in Fig.~\ref{fig:method-descriptions}. 
During learning, we provide our model with training and testing sets generated by 10-fold cross validation. Time series for each feature are processed in parallel using \textit{dynamic time warping} (DTW), which measures the similarity between two time series after finding an optimal match between them by ``warping'' the time axis~\cite{berndt1994using}. This allows the method to absorb some non-linear variations in the time series, for example different speed or resolution of the data. 

For efficiency, we initially apply a time series coarsening strategy called \emph{piece-wise aggregation}. We split each original time series into $p$ equally long sections and replace the time-series values by the section averages, reducing the dimensionality from $L$ to $L'=L/p$. For trend $i$ and feature $k$, we thus obtain a coarsened time series $f^i_k=\{f^i_{k,1}, f^i_{k,2}, \cdots, f^i_{k,L'}\}$. Then, DTW computes the distance between all pairs of points of two given trend time series $f^i_k$ and $f^j_k$. Each element of the resulting $L' \times L'$ distance matrix is $M^{ij}_k(t,t')= (f^i_{k,t} - f^j_{k,t'})^2$. Points closer to each other are more likely to be matched. To create a mapping between the two time series, an optimal path is computed over the time-series distance matrix. A path must start from the beginning of each time series and terminate at its end. The path between first and last points is then computed by minimizing the cumulative distance ($\gamma$) over alternative paths. This problem can be solved via dynamic programming~\cite{berndt1994using} using the following recurrence: $\gamma(t,t') = M(t,t') + min\{\gamma(t-1,t'-1), \gamma(t-1,t'), \gamma(t,t'-1)\}$ (indices $i,j,k$ dropped for readability). The distance $\gamma^{ij}_k$ is used as the $ij$-th element of the $N \times N$ trend similarity matrix $\Gamma_k$.

The computation of similarity between time series using DTW requires $O(L'^2)$ operations. Some heuristic strategies use lower-bounding techniques to reduce the computational complexity~\cite{keogh2005exact}. Another technique is to re-sample the data before adopting DTW. Our coarsening approach reduces the computational costs by a factor of $p^2$. We achieved a significant increase in efficiency with marginal classification accuracy deterioration by setting $p=5$ ($L'=7$). 

In the evaluation step, we use the K-Nearest Neighbor (KNN) algorithm~\cite{cover1967nearest} to assign a class score to a test trend $q$. We compare $q$ with each training trend $i$ to obtain a DTW distance $\gamma^{iq}_k$ for each feature $k$. We then find the $K=5$ labeled trends with smallest DTW distance from $q$, and compute the fraction of promoted trends $s^q_k$ among these nearest neighbors. We finally average across features to obtain the class score $\bar{s^q}$. Higher values of $\bar{s^q}$ indicate a high probability that $q$ is a promoted trend. Class scores, together with ground-truth labels, allow us to computate the AUC of a model, which is then averaged across folds according to cross validation. 

\subsubsection{{\Xsaxvsm} classifier}
\label{sub:sax-vsm}

Our first baseline, called {\Xsaxvsm}, blends symbolic dimensionality reduction and vector space models~\cite{senin2013saxvsm}. 
Time series are encoded via Symbolic Aggregate approXimation (SAX), yielding a compact symbolic representation that has been used for time series anomaly and motif detection, time series clustering, indexing, and more~\cite{lin2003symbolic,lin2007experiencing}. A symbolic representation encodes  numerical features as words. A vector space model is then applied to treat time series as documents for classification purposes, similarly to what is done in information retrieval. 
In our implementation, we first apply piece-wise aggregation and then use SAX to represent the data points in input as a single word of $L'$ letters from an alphabet $\aleph$. This choice and the parameters $|\aleph| = 5$ and $L'=4$ are based on prior optimization~\cite{senin2013saxvsm}, and variations to these settings only marginally affect performance. 
Each time-series value is mapped into a letter by dividing the range of the  feature values into $|\aleph|$ regions in such a way as to obtain equiprobable intervals under an assumption of normality~\cite{lin2007experiencing}. 
In the training phase, for each feature, we build two sets of words corresponding to organic and promoted trends, respectively.  In the test phase, a new instance is assigned to the class with the majority of word matches across features. In case of a tie we assign a random class.
For further details about this baseline and its implementation, we refer the reader to the {\Xsaxvsm} project website.\footnote{\url{github.com/jMotif/sax-vsm_classic}}

\subsubsection{K-Nearest Neighbors classifier}

Our second baseline is an off-the-shelf implementation of the traditional \emph{K-Nearest Neighbors} algorithm~\cite{cover1967nearest} for time-series classification. We used the Python scikit-learn package~\cite{scikit-learn}. 
We selected KNN because it can capture and learn time-series patterns without requiring any pre-processing of the raw time-series data. 
We created the feature vectors for each trend by concatenating into a single vector the continuous-valued time series representing each feature.
The nearest neighbor classifier computes the Euclidean distance between pairs of single-vector time series. For a test trend, the class score is given by the fraction of promoted trends among the $K=5$ nearest neighbors.

\section{Results} 
\label{sec:results}

In this section, we present results of experiments design to evaluate the ability of our machine learning framework to discriminate between organic and promoted trends. 
For all experiments, each feature time series consists of 120 real-valued data points equally divided before and after the trending point. Although in principle we could use the entire time series for classification, ex-post information would not serve our goal of early detection of social media campaigns in a streaming scenario that resembles a real setting, where information about the future evolution of a trend is obviously unavailable.
For this reason, we consider only a subset consisting of $L$ data points ending with delay $D$ since the trending point; 
$D \leq 0$ data points for early detection, $D > 0$ for classification after trending. 
We evaluate the performance of our detection framework as a function of the delay parameter $D$. The case $D=0$ involves detection immediately at trending time. However, we also consider $D < 0$ to examine the performance of our algorithms based on data preceding the trending point; of course the detection would not occur until $D=0$, when one would become aware of the trending hashtag. 
Time series are encoded using the settings described above ($L=35$ windows of length $\ell=6$ hours sliding every $\delta=20$ minutes).

\subsection{Method comparison}

\begin{figure}[!t]
\centering
\includegraphics[width=0.75\columnwidth]{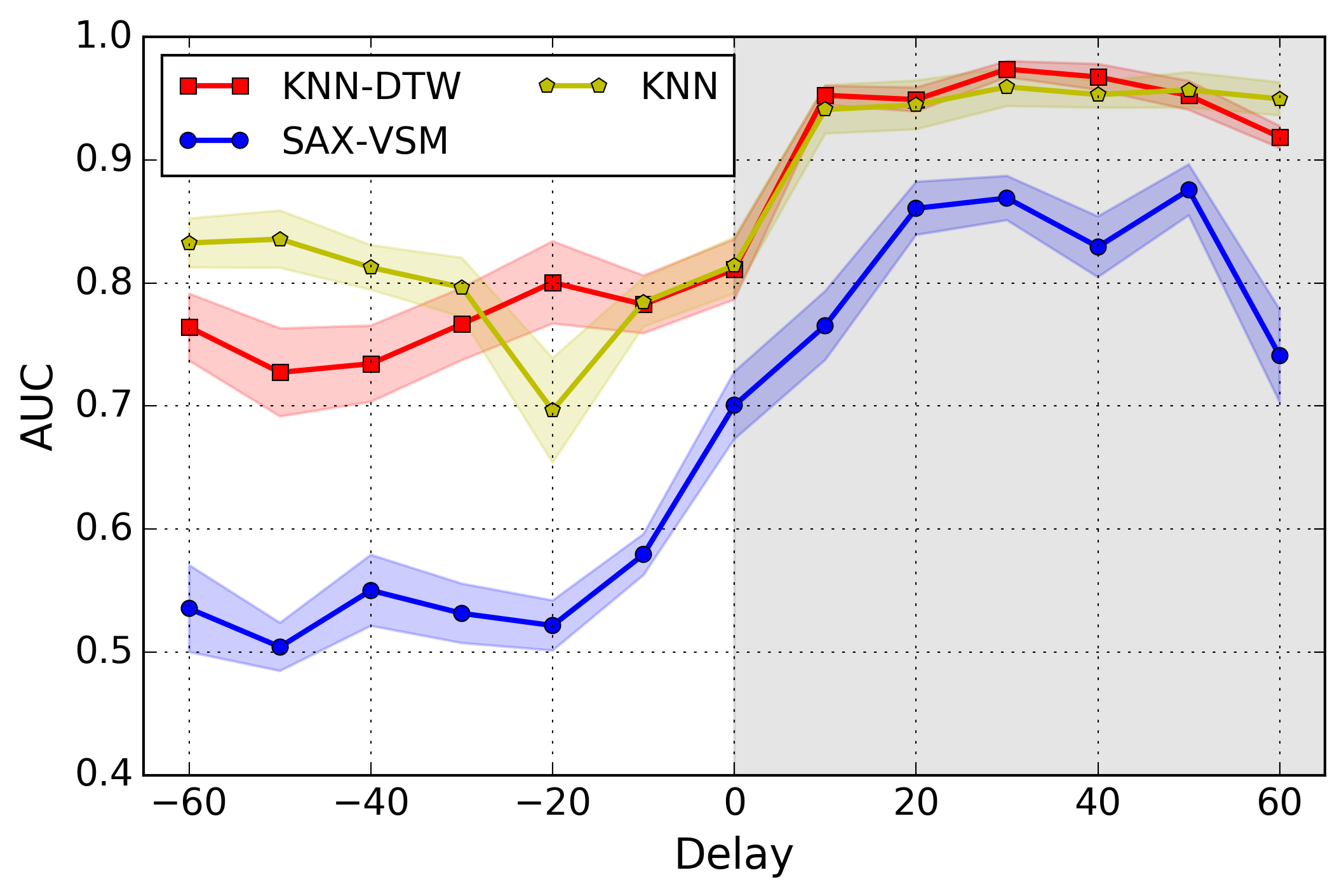}
\caption{Methods comparison. Classification performance of different learning algorithms on encoded and raw time series. The AUC is measured for various delays $D$. Confidence intervals represent standard errors based on 10-fold cross validation.}
\label{fig:method-comparison}
\end{figure}

We carried out an extensive benchmark of several configurations of our system for campaign detection. The performance of the algorithms as a function of varying delays $D$ 
is plotted in Fig.~\ref{fig:method-comparison}. 

\begin{figure}[!t]
\centering
\includegraphics[width=0.75\columnwidth]{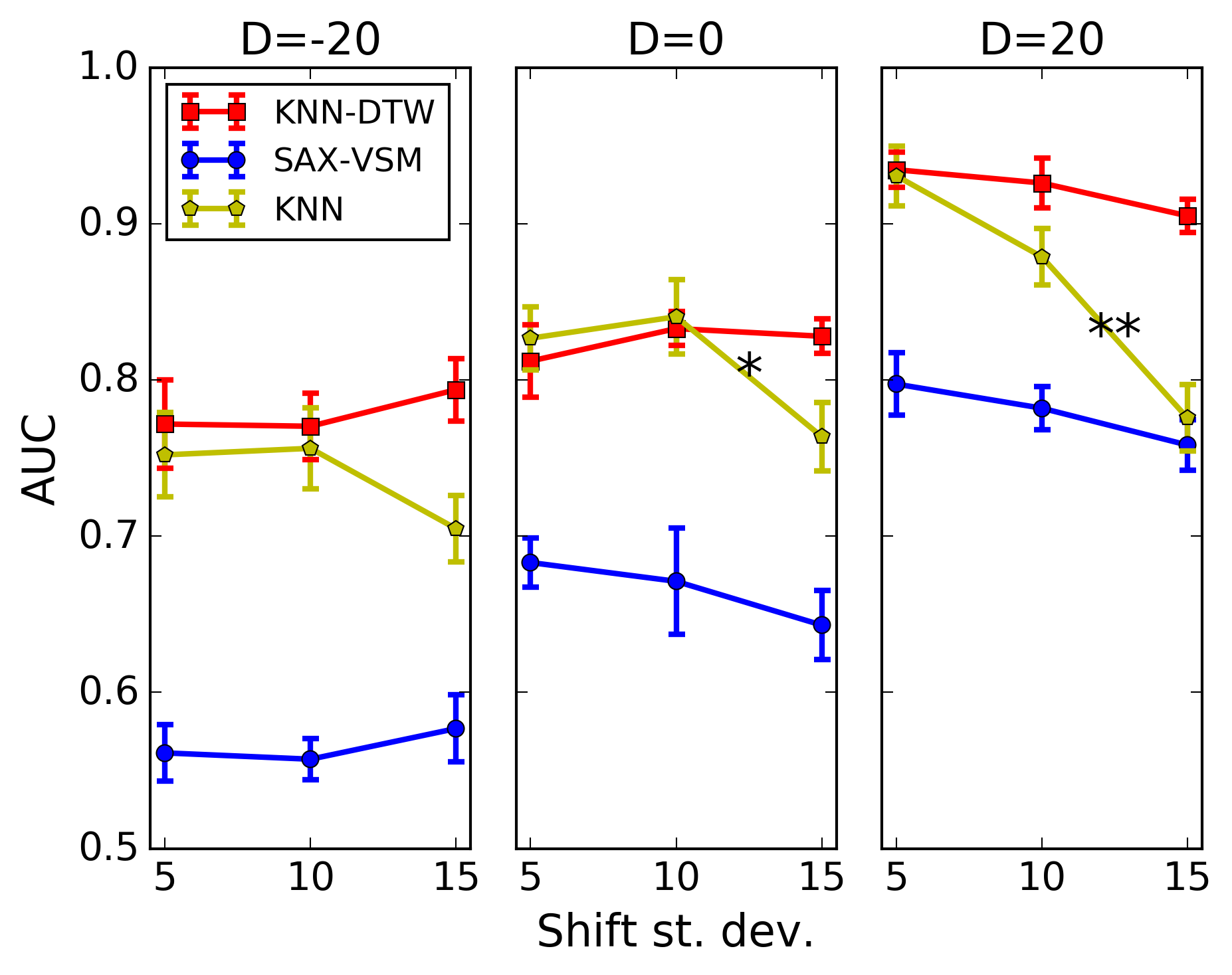}
\caption{Temporal robustness. AUC of different learning algorithms with random temporal shifts versus the standard deviation of the shifts. We repeated the experiment for various delay values $D$. Significance levels of differences in consecutive experiments are marked as (*) $p < 0.05$ and (**) $p<0.01$.}
\label{fig:method-comparison-shifted}
\end{figure}

In addition, we introduce random temporal shifts for each trend time series to test the robustness of the algorithms. In real-world scenarios, one would ideally expect to detect a promoted trend without knowing the trending point. 
To simulate such scenarios, we designed an experiment that introduces variations  that randomly shift each time series around its trending point. The temporal shifts are sampled from gaussian distributions with different variances. We present the results of this experiment in Fig.~\ref{fig:method-comparison-shifted}.

{\Xknndtw} and KNN display the best detection accuracy (measured by AUC) in general. Their performance is comparable (Fig.~\ref{fig:method-comparison}). 
The AUC score is on average around 95\% for detecting promoted trends after trending. In the early detection task, we observe scores above 70\%. 
This is quite remarkable given the small amount of data available before the trending point. {\Xknndtw} also displays a strong robustness to temporal shifts, pointing to the advantage of time warping (Fig.~\ref{fig:method-comparison-shifted}). 
The KNN algorithm is less robust because it computes point-wise similarities between time series without any temporal alignment; as the variance of the temporal shifts increases, we observe a significant drop in accuracy.  
{\Xsaxvsm} benefits from the time series encoding and provides good detection performance (on average around 80\% AUC) but early detection accuracy is poor,  close to random for $D<0$. 
A strong feature of {\Xsaxvsm} is its robustness to temporal shifts, similar to  {\Xknndtw}.

Our experiments suggest that temporal encoding is a crucial ingredient for successful classification of time-series data. Encoding reduces the dimensionality of the signal.  
More importantly, encoding preserves (most) information about temporal trends and makes an algorithm robust to random shifts, which is an importance advantage in real-world scenarios. 
{\Xsaxvsm} ignores long-term temporal ordering. {\Xknndtw}, on the other hand, computes similarities using a time series representation that preserves the long-term temporal order, even as time warping may alter short-term trends. This turns out to be a crucial advantage to achieve both high accuracy and robustness.

\begin{figure} [!t]
\centering
\includegraphics[width=.75\columnwidth]{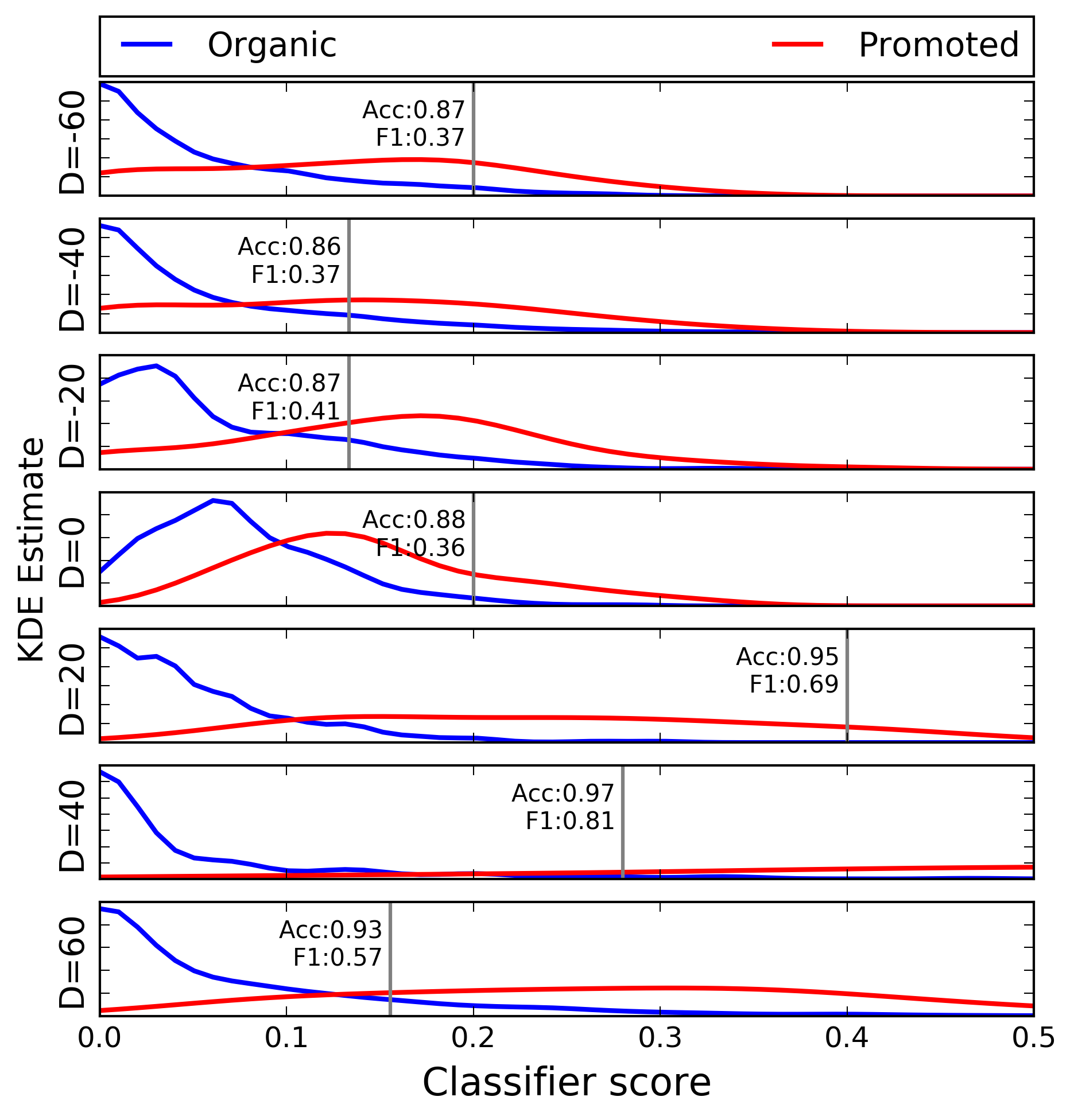}
\caption{Distributions of {\Xknndtw} classifier scores. We use Kernel Density Estimation (KDE), a non-parametric smoothing method, to estimate the probability densities based on finite data samples. 
We also show the threshold values that separate the two classes yielding an optimal F1 score.}
\label{fig:score-distribution}
\end{figure}

Using AUC as an evaluation metric has the advantage of not requiring discretization of scores into binary class labels. However, detection of promoted trends in real scenarios requires binary classification by a threshold. In this way we can measure accuracy, precision, recall, and identify misclassified accounts. 
Fig.~\ref{fig:score-distribution} illustrates the distribution of probabilistic scores produced by the {\Xknndtw} classifier as a function of the delay for the two classes of trends, organic and promoted. 
The scores are computed for leave-out test instances, across folds. An ideal classifier would separate these distributions completely, achieving perfect accuracy. Test instances in the intersection between two distributions either are misclassified or have low-confidence scores. Examples of misclassified instances are discussed in Sec.~\ref{sec:misclassification}.
For $D<0$, {\Xknndtw} generates more conservative scores, and the separation between the organic and promoted class distributions is smaller. 
For $D>0$,  {\Xknndtw} scores separate the two classes well.  To convert continuos scores into binary labels, we calculated the threshold values that maximize the F1 score of each experiment; this score combines precision and recall. Trends with scores above the threshold are labeled as promoted. 
The best accuracy and F1 score are obtained shortly after trending, at $D=20$. 

\subsection{Feature analysis}

\begin{figure} \centering
\includegraphics[width=.75\columnwidth]{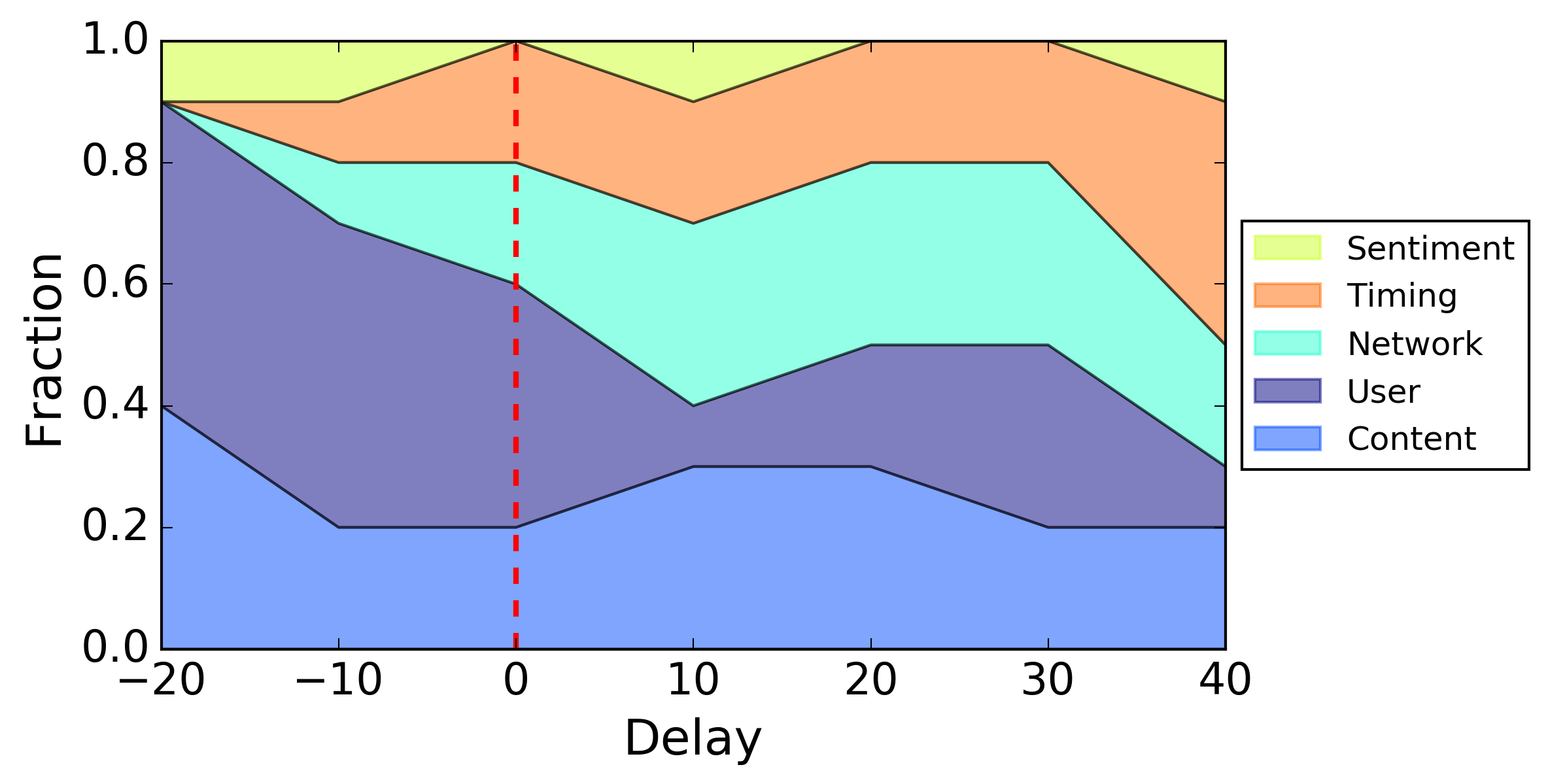}
\caption{{\Xknndtw} feature analysis. Stacked plot showing how different feature classes are represented among the top 10 selected features.}
\label{fig:feature-distribution}
\end{figure}

Let us explore the roles and importance of different features for trend detection. To this end, we identify the significant features using the greedy selection algorithm described in Sec.~\ref{sub:selection}, and group them by the five classes (user meta-data, content, network, sentiment, and timing) previously defined. 
We focus on {\Xknndtw}, our best performing method.
After selecting the top 10 features for different delays $D$, we compute the fractions of top features in each class, as illustrated by Fig.~\ref{fig:feature-distribution}.  
We list the top features for experiments $D=0$ (early detection) and $D=40$ (classification) in Table~\ref{tab:topfeatures}. 

The usefulness of content features does not appear to change significantly between early and late detection.  
In the early detection task, user features seem to contribute significantly more than any other class, possibly because early adopters reveal strong signals about the nature of trends. As we move past the trending point, signals from early adopters are flooded by increasing numbers of participants. Timing and network features become increasingly important as the involvement of more users allows to analyze group activity and network structure patterns. 

\begin{table}[!t]
\caption{Top 10 features for experiments with different values of $D$.}
\begin{tabular}{cll}
\hline
Delay & Features & Classes\\
\hline
40 & Number of tweets & Timing \\
& Max. proportion of pronouns in a tweet & Content \\
& Entropy of hashtag cooccurrence network degree & Network \\
& Entropy of time between two consecutive mentions & Timing \\
& Mean time between two consecutive tweets & Timing \\
& Entropy of emoticon scores & Sentiment \\
& Median time between two consecutive tweets & Timing \\
& Max. originator's followers count & User \\
& Kurtosis of mention network degree distribution & Network \\
& Entropy of pre-determiner POS frequency in a tweet & Content \\
\hline
0 & Max. hashtag cooccurrence network degree & Network \\
& Entropy of number of originator's friends count & User \\
& Max. originator's statuses count & User \\
& Median time between two consecutive tweets & Timing \\
& Skewness of time between two consecutive mentions & Timing \\
& Median of sender's lists count & User \\
& Min. originator's lists count & User \\
& Median of mention network out-degree & Network \\
& Min. frequency of adjective POS in a tweet & Content \\
& Mean frequency of noun POS in a tweet & Content \\
\hline
\end{tabular}
\label{tab:topfeatures}
\end{table}

\subsection{Analysis of misclassifications}
\label{sec:misclassification}

\begin{figure}[!t]
\centering
\includegraphics[width=\columnwidth]{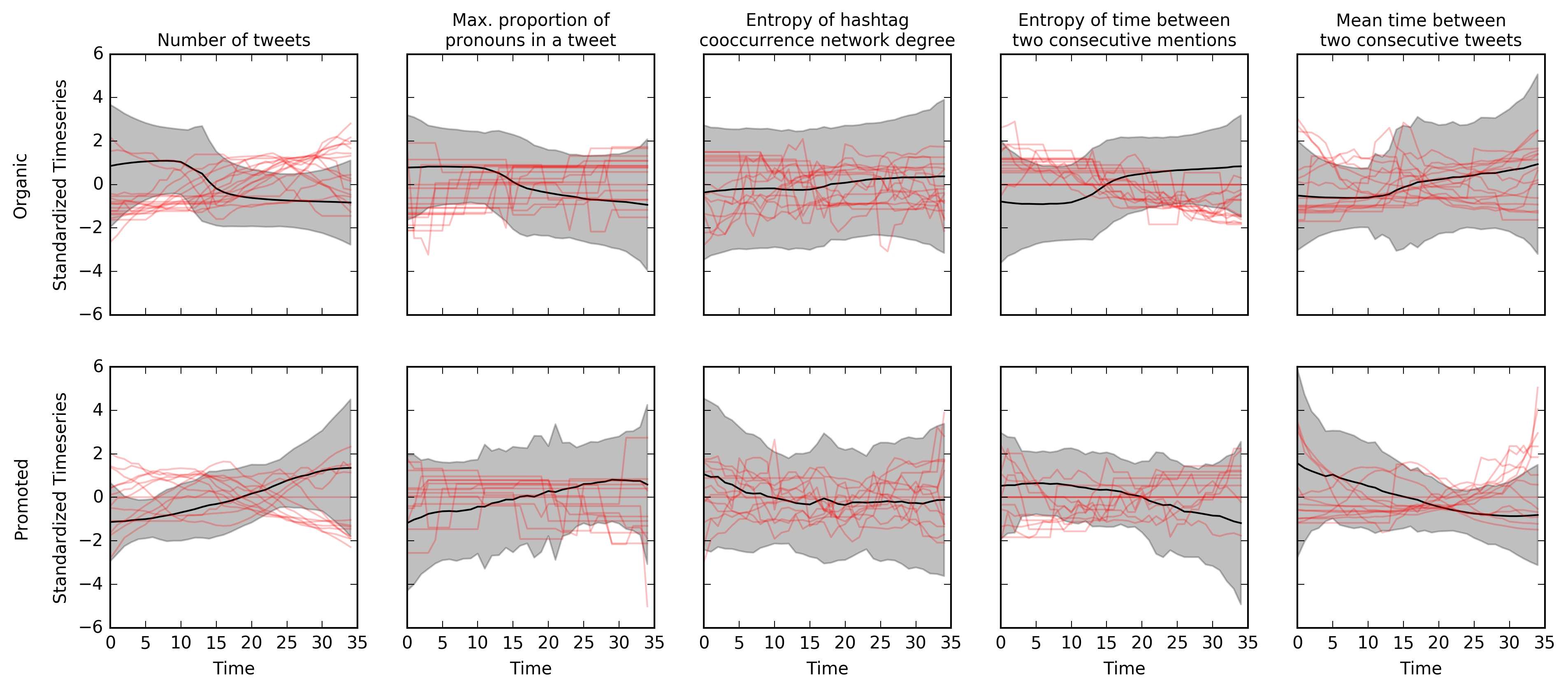}
\caption{Comparison between feature time series of misclassified and correctly classified trends. Time series of the top five features (columns) for promoted (top) and organic (bottom) trends in the $D=40$ detection task. The black lines and gray areas represent the average and 95\% confidence intervals of time series for correctly classified trends. Time series of misclassified trends are shown in red.
Misclassified organic trends (false positives) are: \texttt{\#bobsantigoldlive}, \texttt{\#evildead}, \texttt{\#galaxyfamily}, \texttt{\#gethappy}, \texttt{\#madmen}, \texttt{\#makeboringbrilliant}, \texttt{\#nyias}, \texttt{\#oneboston}, \texttt{\#stingray}, \texttt{\#thewalkingdead}, \texttt{\#timeto365}, \texttt{\#watchsuitstonight}, and \texttt{\#whyiwatchsuits}.
Misclassified promoted trends (false negatives) are: \texttt{\#1dmemories}, \texttt{\#20singersthatilike}, \texttt{\#8thseed}, \texttt{\#bnppo13}, \texttt{\#ciaa}, \texttt{\#expowest}, \texttt{\#jaibrooksforpresident}, \texttt{\#justintimberweek}, \texttt{\#kobalt400}, \texttt{\#mentionsomeonecuteandbeautiful}, \texttt{\#nyc}, \texttt{\#realestate}, \texttt{\#stars}, \texttt{\#sxsw}, \texttt{\#wbc}, and \texttt{\#wcw}.}
\label{fig:feature-timeseries-comparison}
\end{figure}

We conclude our analysis by discussing when our system fails. 
In Fig.~\ref{fig:feature-timeseries-comparison}, we illustrate how some key features of misclassified trends diverge from the majority of the trends that are correctly classified.  We observe that some misclassified trends follow the temporal characteristics of the other class. This is best illustrated in the case of volume (number of tweets). 

An advantage of continuous class scores is that we can tune the classification threshold to achieve a desired balance between precision and recall, or between false positives and false negatives.
False negative errors are the most costly for a detection system: a promoted trend mistakenly labeled as organic would easily go unchecked among the larger number of correctly labeled organic trends.
Focusing our attention on a few specific instances of false negatives generated by our system, we gained some insight on the reasons triggering the mistakes.
First of all, it is conceivable that promoted trends are sustained by organic activity before promotion and therefore they are essentially indistinguishable from organic ones until the promotion triggers the trending behavior. 
It is also reasonable to expect a decline in performance for long delays: as more users join the conversation, promoted trends become harder to distinguish from organic ones. This may explain the dip in accuracy observed for the longest delay (cf. Fig.~\ref{fig:method-comparison}). 

False positives (organic trends mistakenly labeled as promoted) can be manually filtered out in post-processing and are therefore less costly. However, analysis of false positives provides for some insight as well.  
Some trends in our dataset, such as \texttt{\#watchsuitstonight} and \texttt{\#madmen}, were promoted via alternative communication channels (television and radio), rather than via Twitter. 
This has become a common practice in recent years, as more and more Twitter campaigns are mentioned or advertised externally to trigger organic-looking responses in the audience. 
Our system recognized such instances as promoted, whereas their ground-truth labels did not.  Those campaigns were therefore wrongly counted as false positives, penalizing our algorithms in the evaluation.
We find it remarkable that in these cases our system is capable of learning the signature of promoted trends, even though the promotion occurs outside of the social media itself.

\section{Related work} 
\label{sec:related-work}

Recent work on social media provides a better understanding of human communication dynamics such as collective attention and information diffusion \cite{weng2012competition}, the emergence of trends \cite{leskovec2009meme,ferrara2013traveling}, social influence and political mobilization \cite{bond201261,conover2013geospatial,conover2013digital,varol2014evolution}.

Different information diffusion mechanisms may determine the trending dynamics of hashtags and other memes on social media. Exogenous and endogenous dynamics produce memes with distinctive characteristics~\cite{sornette2004endogenous,myers2012information,lehmann2012dynamical,ferrara2013traveling,ferrara2015quantifying}: external events occurring in the real world (e.g., a natural disaster or a terrorist attack) can generate chatter on the platform and therefore trigger the trending of a new, unforeseen hashtag; other topics (e.g., politics or entertainment) are continuously discussed and sometimes a particular conversation can accrue lots of attention and generate trending memes. The promotional campaigns studied here can be seen as a type of exogenous factor affecting the visibility of memes.

The present work, to the best of our knowledge, is the first to investigate the early detection of promoted content on social media.
We focus our attention on advertisement, which can play an important role in information campaigns.
Trending memes are considered an indicator of collective attention in social media \cite{wu2007novelty,lehmann2012dynamical}, and as such they have been used to predict real-world events, like the winner of a popular reality TV show \cite{ciulla2012beating}.
Although emerging from collective attention, communication on social media can be manipulated, for example for political gain, as in the case of astroturf \cite{metaxas2012social,ratkiewicz2011detecting}. 

Recent work analyzes emerging topics, memes, and conversations triggered by real world events~\cite{aggarwal2012event,becker2011beyond,cataldi2010emerging}.
Studies of information dissemination reveal mechanisms governing content production and consumption~\cite{Ciampaglia2015production} as well as prediction of future content popularity. 
Cheng \textit{et al.} study the prediction of photo-sharing cascade size~\cite{cheng2014can} and recurrence~\cite{cheng2016cascades} on Facebook. Machine learning models can predict future popularity of emerging hashtags and content on social media~\cite{tsur2012s,ma2013predicting}. Features extracted from content~\cite{jamali2009digging}, sentiment~\cite{ferrara2015quantifying,krauss2008predicting}, community structure~\cite{weng2013virality,weng2013role}, and temporal signatures~\cite{pinto2013using,figueiredo2011tube,wang2015burst} are commonly used to train such models.
In this paper we leverage similar features, but for the novel task of campaign detection. Furthermore, our task is more challenging because we deal with dynamic features whose changes over time are captured in high-dimensional time series.

Another topic related to our research is rumor detection. Rumors may  emerge organically as genuine conversation and spread out of control. They are characterized and sustained by ambiguous contexts, where correctness and completeness of information or the meaning of a situation is not obviously apparent \cite{difonzo2007}. 
Examples are situations of crisis or topics of public debate \cite{mendoza2010twitter}.
Existing systems to identify rumors are based mostly on content analysis \cite{qazvinian2011rumor,kwon2013prominent} and clustering techniques~\cite{ferrara2013clustering,jafariasbagh2014clustering}. An open question is to determine if rumor detection might benefit from the wide set of feature classes we propose here.

The proposed framework is based on a mixture of features common in social media data, including emotional and sentiment information.
The literature has reported extensively on the use of social media content to describe emotional and demographic characteristics of users \cite{mislove2011understanding,ferrara2015quantifying,mitchell2013geography}.
The use of language in online communities is the focus of two recent papers  \cite{danescu2013no,mcauley2013amateurs}: the authors observe that the language of social media users evolves, and common patterns emerge over time. 
The language style of users adapts to achieve better fitness in the conversation~\cite{das2016information}.
These findings suggest that language contains strong signals, in particular if studied in conjunction with other dimensions of the data. Our study confirms the importance of content for campaign detection.

Finally, our system builds on network features and diffusion patterns of social media messages. 
Network structure and information diffusion in social media have been studied extensively \cite{backstrom2006group,lerman2010information}.
Network features are highly predictive of certain types of social media abuse, like astroturf, that attempt to simulate grassroots online conversations  \cite{ratkiewicz2011detecting,ratkiewicz2011truthy,subrahmanian2016darpa,ferrara2016rise,varol2017online}. 
Such artificial campaigns produce peculiar patterns of information diffusion: the topology of retweet or mention networks is often a stronger signal than content or language. The present findings are consistent with this body of work, as network features are helpful in detecting promoted content after trending.

\section{Conclusions} 
\label{sec:conclusions}

As we increasingly rely on social media to satisfy our information needs, it is important to recognize the dynamics behind online campaigns.
In this paper, we posed the problem of early-detection of promoted trends on social media, discussed the challenges that this problem presents, 
and proposed a supervised computational framework to attack it. 
The proposed system leverages time series representing the evolution of different features characterizing trending campaigns. The list includes  features relative to network structure and diffusion patterns, sentiment, language and content features, timing, and user meta-data.
We demonstrated the crucial advantages of encoding temporal sequences. 

We achieved good accuracy in campaign detection. Our early detection performance is  remarkable when one considers the challenging nature of the problem and the low volume of data available in the early stage of a campaign.
We also studied the robustness of the proposed algorithms by introducing random temporal shifts around the trending point, simulating realistic scenarios in which the trending point can only be estimated with limited accuracy.

One of the advantages of our framework is that of providing interpretable feature classes. 
We explored how content, network, and user features affect detection performance. 
Extensive feature analysis revealed that signatures of campaigns can be detected early, especially by leveraging content and user features. After the trending point, network and temporal features become more useful.

The availability of data about organic and promoted trends is subject to Twitter's recipe for selecting trending hashtags. There is no certain way to know if and when social media platforms make any changes to such recipes. However, nothing in our approach assumes any knowledge of a particular platform's trending recipe. If the recipe changes, our system could be retrained accordingly. 

This work represents an important step toward the automatic detection of campaigns. 
The problem is of paramount importance, since social media shape the opinions of millions of users in everyday life.
Further work is needed to study whether different classes of campaigns (say, legitimate advertising vs. terrorist propaganda) may exhibit characteristics captured by distinct features. 
Many of the features leveraged in our model, such as those related to network structure and temporal attributes, capture activity patterns that could provide useful signals to detect astroturf \cite{ratkiewicz2011detecting}.  Therefore, our framework could in principle be applied to astroturf detection, if longitudinal training data about astroturf campaigns were available.

\section*{Acknowledgements}
We thank Mohsen JafariAsbagh, Qiaozhu Mei, Zhe Zhao, and Sergey Malinchik for helpful discussions in 2012 and 2013. We are also grateful to two anonymous reviewers, whose suggestions greatly improved this paper. This work was supported by ONR (N15A-020-0053), NSF (grant CCF-1101743), DARPA (grant W911NF-12-1-0037), and the McDonnell Foundation. The funders had no role in study design, data collection and analysis, decision to publish, or preparation of the manuscript.

\bibliographystyle{plain}
\bibliography{references}

\begin{thebibliography}{10}

\bibitem{agarwal2011sentiment}
Apoorv Agarwal, Boyi Xie, Ilia Vovsha, Owen Rambow, and Rebecca Passonneau.
\newblock Sentiment analysis of {T}witter data.
\newblock In {\em Proceedings of the Workshop on Languages in Social Media},
  pages 30--38. ACL, 2011.

\bibitem{aggarwal2012event}
Charu~C Aggarwal and Karthik Subbian.
\newblock Event detection in social streams.
\newblock In {\em SDM}, volume~12, pages 624--635. SIAM, 2012.

\bibitem{backstrom2006group}
Lars Backstrom, Dan Huttenlocher, Jon Kleinberg, and Xiangyang Lan.
\newblock Group formation in large social networks: membership, growth, and
  evolution.
\newblock In {\em Proc. of the 12th ACM SIGKDD international conference on
  Knowledge discovery and data mining}, pages 44--54, 2006.

\bibitem{bakshy2011everyone}
Eytan Bakshy, Jake~M Hofman, Winter~A Mason, and Duncan~J Watts.
\newblock Everyone's an influencer: quantifying influence on {T}witter.
\newblock In {\em Proc. of the 4th ACM international conference on Web search
  and data mining}, pages 65--74, 2011.

\bibitem{becker2011beyond}
Hila Becker, Mor Naaman, and Luis Gravano.
\newblock Beyond trending topics: Real-world event identification on twitter.
\newblock {\em ICWSM}, 11:438--441, 2011.

\bibitem{berger2015isis}
J~Berger and Jonathan Morgan.
\newblock The isis twitter census: Defining and describing the population of
  isis supporters on twitter.
\newblock {\em The Brookings Project on US Relations with the Islamic World},
  3:20, 2015.

\bibitem{berndt1994using}
Donald~J Berndt and James Clifford.
\newblock Using dynamic time warping to find patterns in time series.
\newblock In {\em Proc. of AAAI Workshop on Knowledge Discovery in Databases},
  pages 359--370. Seattle, WA, 1994.

\bibitem{bessi2015science}
Alessandro Bessi, Mauro Coletto, George~Alexandru Davidescu, Antonio Scala,
  Guido Caldarelli, and Walter Quattrociocchi.
\newblock Science vs conspiracy: Collective narratives in the age of
  misinformation.
\newblock {\em PLoS ONE}, 10(2):e0118093, 02 2015.

\bibitem{bessi2016social}
Alessandro Bessi and Emilio Ferrara.
\newblock Social bots distort the 2016 us presidential election online
  discussion.
\newblock {\em First Monday}, 21(11), 2016.

\bibitem{bollen2011twitter}
Johan Bollen, Huina Mao, and Xiaojun Zeng.
\newblock {T}witter mood predicts the stock market.
\newblock {\em Journal of Computational Science}, 2(1):1--8, 2011.

\bibitem{bond201261}
Robert~M Bond, Christopher~J Fariss, Jason~J Jones, Adam~DI Kramer, Cameron
  Marlow, Jaime~E Settle, and James~H Fowler.
\newblock A 61-million-person experiment in social influence and political
  mobilization.
\newblock {\em Nature}, 489(7415):295--298, 2012.

\bibitem{botta2015quantifying}
Federico Botta, Helen~Susannah Moat, and Tobias Preis.
\newblock Quantifying crowd size with mobile phone and twitter data.
\newblock {\em Royal Society open science}, 2(5):150162, 2015.

\bibitem{briscoe2014cues}
E~Briscoe, S~Appling, and H~Hayes.
\newblock Cues to deception in social media communications.
\newblock In {\em Proceedings of the Hawaii International Conference on System
  Sciences}, 2014.

\bibitem{cataldi2010emerging}
Mario Cataldi, Luigi Di~Caro, and Claudio Schifanella.
\newblock Emerging topic detection on twitter based on temporal and social
  terms evaluation.
\newblock In {\em Proceedings of the Tenth International Workshop on Multimedia
  Data Mining}, page~4. ACM, 2010.

\bibitem{cheng2014can}
Justin Cheng, Lada Adamic, P~Alex Dow, Jon~Michael Kleinberg, and Jure
  Leskovec.
\newblock Can cascades be predicted?
\newblock In {\em Proceedings of the 23rd international conference on World
  wide web}, pages 925--936. ACM, 2014.

\bibitem{cheng2016cascades}
Justin Cheng, Lada~A Adamic, Jon~M Kleinberg, and Jure Leskovec.
\newblock Do cascades recur?
\newblock In {\em Proceedings of the 25th International Conference on World
  Wide Web}, pages 671--681. International World Wide Web Conferences Steering
  Committee, 2016.

\bibitem{Ciampaglia2015production}
Giovanni~Luca Ciampaglia, Alessandro Flammini, and Filippo Menczer.
\newblock The production of information in the attention economy.
\newblock {\em Scientific Reports}, 5:9452, 2015.

\bibitem{ciampaglia2015computational}
Giovanni~Luca Ciampaglia, Prashant Shiralkar, Luis~M. Rocha, Johan Bollen,
  Filippo Menczer, and Alessandro Flammini.
\newblock Computational fact checking from knowledge networks.
\newblock {\em PLoS ONE}, 10(6):e0128193, 06 2015.

\bibitem{ciulla2012beating}
Fabio Ciulla, Delia Mocanu, Andrea Baronchelli, Bruno Gon{\c{c}}alves, Nicola
  Perra, and Alessandro Vespignani.
\newblock Beating the news using social media: the case study of {A}merican
  {I}dol.
\newblock {\em EPJ Data Science}, 1(1):1--11, 2012.

\bibitem{clark2015vaporous}
Eric~M Clark, Chris~A Jones, Jake~Ryland Williams, Allison~N Kurti,
  Michell~Craig Nortotsky, Christopher~M Danforth, and Peter~Sheridan Dodds.
\newblock Vaporous marketing: Uncovering pervasive electronic cigarette
  advertisements on twitter.
\newblock {\em arXiv preprint arXiv:1508.01843}, 2015.

\bibitem{conover2013geospatial}
Michael~D Conover, Clayton Davis, Emilio Ferrara, Karissa McKelvey, Filippo
  Menczer, and Alessandro Flammini.
\newblock The geospatial characteristics of a social movement communication
  network.
\newblock {\em PloS ONE}, 8:e55957, 2013.

\bibitem{conover2013digital}
Michael~D Conover, Emilio Ferrara, Filippo Menczer, and Alessandro Flammini.
\newblock The digital evolution of {O}ccupy {W}all {S}treet.
\newblock {\em PloS ONE}, 8:e64679, 2013.

\bibitem{cover1967nearest}
Thomas~M Cover and Peter~E Hart.
\newblock Nearest neighbor pattern classification.
\newblock {\em Information Theory, IEEE Transactions on}, 13(1):21--27, 1967.

\bibitem{danescu2013no}
Cristian Danescu-Niculescu-Mizil, Robert West, Dan Jurafsky, Jure Leskovec, and
  Christopher Potts.
\newblock No country for old members: user lifecycle and linguistic change in
  online communities.
\newblock In {\em Proceedings of the 22nd international conference on World
  Wide Web}, pages 307--318, 2013.

\bibitem{das2016information}
Abhimanyu Das, Sreenivas Gollapudi, Emre K{\i}c{\i}man, and Onur Varol.
\newblock Information dissemination in heterogeneous-intent networks.
\newblock In {\em Proceedings of the 8th ACM Conference on Web Science}, pages
  259--268. ACM, 2016.

\bibitem{davis2016botornot}
Clayton~Allen Davis, Onur Varol, Emilio Ferrara, Alessandro Flammini, and
  Filippo Menczer.
\newblock Botornot: A system to evaluate social bots.
\newblock In {\em Proceedings of the 25th International Conference Companion on
  World Wide Web}, pages 273--274. International World Wide Web Conferences
  Steering Committee, 2016.

\bibitem{difonzo2007}
Nicholas DiFonzo and Prashant Bordia.
\newblock Rumor, gossip and urban legends.
\newblock {\em Diogenes}, 54(1):19--35, 2007.

\bibitem{fawcett2006introduction}
Tom Fawcett.
\newblock {An introduction to ROC analysis}.
\newblock {\em Pattern recognition letters}, 27(8):861--874, 2006.

\bibitem{ferrara2013clustering}
Emilio Ferrara, Mohsen JafariAsbagh, Onur Varol, Vaheda Qazvinian, Filippo
  Menczer, and Alessandro Flammini.
\newblock Clustering memes in social media.
\newblock In {\em Proceedings of the 2013 IEEE/ACM International Conference on
  Advances in Social Networks Analysis and Mining}, pages 548--555. IEEE/ACM,
  2013.

\bibitem{ferrara2016rise}
Emilio Ferrara, Onur Varol, Clayton Davis, Filippo Menczer, and Alessandro
  Flammini.
\newblock The rise of social bots.
\newblock {\em Communications of the ACM}, 59(7):96--104, 2016.

\bibitem{ferrara2013traveling}
Emilio Ferrara, Onur Varol, Filippo Menczer, and Alessandro Flammini.
\newblock Traveling trends: social butterflies or frequent fliers?
\newblock In {\em Proc. of the first ACM conference on Online social networks},
  pages 213--222, 2013.

\bibitem{ferrara2016predicting}
Emilio Ferrara, Wen-Qiang Wang, Onur Varol, Alessandro Flammini, and Aram
  Galstyan.
\newblock Predicting online extremism, content adopters, and interaction
  reciprocity.
\newblock In {\em International Conference on Social Informatics}, pages
  22--39. Springer, 2016.

\bibitem{ferrara2015quantifying}
Emilio Ferrara and Zeyao Yang.
\newblock Quantifying the effect of sentiment on information diffusion in
  social media.
\newblock {\em PeerJ Computer Science}, 1:e26, 2015.

\bibitem{figueiredo2011tube}
Flavio Figueiredo, Fabr{\'\i}cio Benevenuto, and Jussara~M Almeida.
\newblock The tube over time: characterizing popularity growth of youtube
  videos.
\newblock In {\em Proceedings of the fourth ACM international conference on Web
  search and data mining}, pages 745--754. ACM, 2011.

\bibitem{Ghosh11snakdd}
Rumi Ghosh, Tawan Surachawala, and Kristina Lerman.
\newblock Entropy-based classification of retweeting activity on twitter.
\newblock In {\em Proceedings of KDD workshop on Social Network Analysis
  (SNA-KDD)}, August 2011.

\bibitem{guyon2003introduction}
Isabelle Guyon and Andr{\'e} Elisseeff.
\newblock An introduction to variable and feature selection.
\newblock {\em The Journal of Machine Learning Research}, 3:1157--1182, 2003.

\bibitem{haustein2016tweets}
Stefanie Haustein, Timothy~D Bowman, Kim Holmberg, Andrew Tsou, Cassidy~R
  Sugimoto, and Vincent Larivi{\`e}re.
\newblock Tweets as impact indicators: Examining the implications of automated
  ``bot'' accounts on twitter.
\newblock {\em Journal of the Association for Information Science and
  Technology}, 67(1):232--238, 2016.

\bibitem{jafariasbagh2014clustering}
Mohsen JafariAsbagh, Emilio Ferrara, Onur Varol, Filippo Menczer, and
  Alessandro Flammini.
\newblock Clustering memes in social media streams.
\newblock {\em Social Network Analysis and Mining}, 4(1):1--13, 2014.

\bibitem{jamali2009digging}
Salman Jamali and Huzefa Rangwala.
\newblock Digging digg: Comment mining, popularity prediction, and social
  network analysis.
\newblock In {\em Web Information Systems and Mining, 2009. WISM 2009.
  International Conference on}, pages 32--38. IEEE, 2009.

\bibitem{john1994irrelevant}
George~H John, Ron Kohavi, Karl Pfleger, et~al.
\newblock Irrelevant features and the subset selection problem.
\newblock In {\em Machine learning: proceedings of the eleventh international
  conference}, pages 121--129, 1994.

\bibitem{keogh2005exact}
Eamonn Keogh and Chotirat~Ann Ratanamahatana.
\newblock Exact indexing of dynamic time warping.
\newblock {\em Knowledge and information systems}, 7(3):358--386, 2005.

\bibitem{kloumann2012positivity}
Isabel~M Kloumann, Christopher~M Danforth, Kameron~Decker Harris, Catherine~A
  Bliss, and Peter~Sheridan Dodds.
\newblock Positivity of the english language.
\newblock {\em PloS one}, 7(1):e29484, 2012.

\bibitem{krauss2008predicting}
Jonas Krauss, Stefan Nann, Daniel Simon, Peter~A Gloor, and Kai Fischbach.
\newblock Predicting movie success and academy awards through sentiment and
  social network analysis.
\newblock In {\em ECIS}, pages 2026--2037, 2008.

\bibitem{kwon2013prominent}
Sejeong Kwon, Meeyoung Cha, Kyomin Jung, Wei Chen, and Yajun Wang.
\newblock Prominent features of rumor propagation in online social media.
\newblock In {\em Proc. IEEE International Conference on Data Mining series
  (ICDM)}, 2013.

\bibitem{lehmann2012dynamical}
Janette Lehmann, Bruno Gon{\c{c}}alves, Jos{\'e}~J Ramasco, and Ciro Cattuto.
\newblock Dynamical classes of collective attention in {T}witter.
\newblock In {\em Proc. the 21th International Conference on World Wide Web},
  pages 251--260, 2012.

\bibitem{lerman2010information}
Kristina Lerman and Rumi Ghosh.
\newblock Information contagion: An empirical study of the spread of news on
  {D}igg and {T}witter social networks.
\newblock In {\em Proceedings of the 4th International AAAI Conference on
  Weblogs and Social Media}, pages 90--97, 2010.

\bibitem{leskovec2009meme}
Jure Leskovec, Lars Backstrom, and Jon Kleinberg.
\newblock Meme-tracking and the dynamics of the news cycle.
\newblock In {\em Proceedings of the 15th ACM SIGKDD International Conference
  on Knowledge Discovery and Data Mining}, pages 497--506. ACM, 2009.

\bibitem{letchford2015advantage}
Adrian Letchford, Helen~Susannah Moat, and Tobias Preis.
\newblock The advantage of short paper titles.
\newblock {\em Royal Society Open Science}, 2(8):150266, 2015.

\bibitem{lin2003symbolic}
Jessica Lin, Eamonn Keogh, Stefano Lonardi, and Bill Chiu.
\newblock A symbolic representation of time series, with implications for
  streaming algorithms.
\newblock In {\em Proceedings of the 8th ACM SIGMOD workshop on Research issues
  in data mining and knowledge discovery}, pages 2--11, 2003.

\bibitem{lin2007experiencing}
Jessica Lin, Eamonn Keogh, Li~Wei, and Stefano Lonardi.
\newblock Experiencing sax: a novel symbolic representation of time series.
\newblock {\em Data Mining and Knowledge Discovery}, 15(2):107--144, 2007.

\bibitem{ma2013predicting}
Zongyang Ma, Aixin Sun, and Gao Cong.
\newblock On predicting the popularity of newly emerging hashtags in twitter.
\newblock {\em Journal of the American Society for Information Science and
  Technology}, 64(7):1399--1410, 2013.

\bibitem{mcauley2013amateurs}
Julian~John McAuley and Jure Leskovec.
\newblock From amateurs to connoisseurs: modeling the evolution of user
  expertise through online reviews.
\newblock In {\em Proceedings of the 22nd international conference on World
  Wide Web}, pages 897--908. ACM, 2013.

\bibitem{mendoza2010twitter}
Marcelo Mendoza, Barbara Poblete, and Carlos Castillo.
\newblock {T}witter under crisis: Can we trust what we {RT}?
\newblock In {\em Proceedings of the first workshop on social media analytics},
  pages 71--79. ACM, 2010.

\bibitem{metaxas2012social}
Panagiotis~T Metaxas and Eni Mustafaraj.
\newblock Social media and the elections.
\newblock {\em Science}, 338(6106):472--473, 2012.

\bibitem{mislove2011understanding}
Alan Mislove, Sune Lehmann, Yong-Yeol Ahn, Jukka-Pekka Onnela, and J~Niels
  Rosenquist.
\newblock Understanding the demographics of {T}witter users.
\newblock In {\em Proceedings of the 5th International AAAI Conference on
  Weblogs and Social Media}, 2011.

\bibitem{mitchell2013geography}
Lewis Mitchell, Kameron~Decker Harris, Morgan~R Frank, Peter~Sheridan Dodds,
  and Christopher~M Danforth.
\newblock The geography of happiness: Connecting {T}witter sentiment and
  expression, demographics, and objective characteristics of place.
\newblock {\em PloS one}, 8(5):e64417, 2013.

\bibitem{mocanu2013twitter}
Delia Mocanu, Andrea Baronchelli, Nicola Perra, Bruno Gon\c{c}alves, Qian
  Zhang, and Alessandro Vespignani.
\newblock The {T}witter of {B}abel: Mapping world languages through
  microblogging platforms.
\newblock {\em PloS one}, 8(4):e61981, January 2013.

\bibitem{myers2014bursty}
Seth~A Myers and Jure Leskovec.
\newblock The bursty dynamics of the twitter information network.
\newblock In {\em Proceedings of the 23rd international conference on World
  wide web}, pages 913--924. ACM, 2014.

\bibitem{myers2012information}
Seth~A Myers, Chenguang Zhu, and Jure Leskovec.
\newblock Information diffusion and external influence in networks.
\newblock In {\em Proceedings of the 18th ACM SIGKDD international conference
  on Knowledge discovery and data mining}, pages 33--41. ACM, 2012.

\bibitem{olteanu2017distilling}
Alexandra Olteanu, Onur Varol, and Emre K{\i}c{\i}man.
\newblock Distilling the outcomes of personal experiences: A propensity-scored
  analysis of social media.
\newblock In {\em Proc. of The 20th ACM Conference on Computer-Supported
  Cooperative Work and Social Computing}, 2017.

\bibitem{scikit-learn}
F.~Pedregosa, G.~Varoquaux, A.~Gramfort, V.~Michel, B.~Thirion, O.~Grisel,
  M.~Blondel, P.~Prettenhofer, R.~Weiss, V.~Dubourg, J.~Vanderplas, A.~Passos,
  D.~Cournapeau, M.~Brucher, M.~Perrot, and E.~Duchesnay.
\newblock Scikit-learn: Machine learning in {P}ython.
\newblock {\em Journal of Machine Learning Research}, 12:2825--2830, 2011.

\bibitem{pinto2013using}
Henrique Pinto, Jussara~M Almeida, and Marcos~A Gon{\c{c}}alves.
\newblock Using early view patterns to predict the popularity of youtube
  videos.
\newblock In {\em Proceedings of the sixth ACM international conference on Web
  search and data mining}, pages 365--374. ACM, 2013.

\bibitem{qazvinian2011rumor}
Vahed Qazvinian, Emily Rosengren, Dragomir~R Radev, and Qiaozhu Mei.
\newblock Rumor has it: Identifying misinformation in microblogs.
\newblock In {\em Proceedings of the Conference on Empirical Methods in Natural
  Language Processing}, pages 1589--1599. ACL, 2011.

\bibitem{ratkiewicz2011detecting}
J.~Ratkiewicz, M.~Conover, M.~Meiss, B.~Goncalves, A.~Flammini, and F.~Menczer.
\newblock Detecting and tracking political abuse in social media.
\newblock In {\em Proceedings of the 5th International AAAI Conference on
  Weblogs and Social Media}, pages 297--304, 2011.

\bibitem{ratkiewicz2011truthy}
J.~Ratkiewicz, M.~Conover, M.~Meiss, B.~Gon{\c{c}}alves, S.~Patil, A.~Flammini,
  and F.~Menczer.
\newblock Truthy: mapping the spread of astroturf in microblog streams.
\newblock In {\em Proceedings of the 20th International Conference on World
  Wide Web}, pages 249--252, 2011.

\bibitem{senin2013saxvsm}
Pavel Senin and Sergey Malinchik.
\newblock Sax-vsm: Interpretable time series classification using sax and
  vector space model.
\newblock In {\em Data Mining (ICDM), 2013 IEEE 13th International Conference
  on}, pages 1175--1180. IEEE, 2013.

\bibitem{guardian2015}
Maeve Shearlaw.
\newblock From britain to beijing: how governments manipulate the internet.
\newblock Accessed online at
  http://www.theguardian.com/world/2015/apr/02/russia-troll-factory-kremlin-cyber-army-comparisons,
  April 2015.

\bibitem{sornette2004endogenous}
Didier Sornette, Fabrice Desch{\^a}tres, Thomas Gilbert, and Yann Ageon.
\newblock Endogenous versus exogenous shocks in complex networks: An empirical
  test using book sale rankings.
\newblock {\em Physical Review Letters}, 93(22):228701, 2004.

\bibitem{subrahmanian2016darpa}
VS~Subrahmanian, Amos Azaria, Skylar Durst, Vadim Kagan, Aram Galstyan,
  Kristina Lerman, Linhong Zhu, Emilio Ferrara, Alessandro Flammini, Filippo
  Menczer, Andrew Stevens, Alexander Dekhtyar, Shuyang Gao, Tad Hogg, Farshad
  Kooti, Yan Liu, Onur Varol, Prashant Shiralkar, Vinod Vydiswaran, Qiaozhu
  Mei, and Tim Hwang.
\newblock {The DARPA Twitter Bot Challenge}.
\newblock {\em IEEE Computer}, 49(6):38--46, 2016.
\newblock Preprint arXiv:1601.05140.

\bibitem{tsur2012s}
Oren Tsur and Ari Rappoport.
\newblock What's in a hashtag?: content based prediction of the spread of ideas
  in microblogging communities.
\newblock In {\em Proceedings of the fifth ACM international conference on Web
  search and data mining}, pages 643--652. ACM, 2012.

\bibitem{tumasjan2010predicting}
Andranik Tumasjan, Timm~Oliver Sprenger, Philipp~G Sandner, and Isabell~M
  Welpe.
\newblock Predicting elections with twitter: What 140 characters reveal about
  political sentiment.
\newblock {\em ICWSM}, 10:178--185, 2010.

\bibitem{twittertrend}
{Twitter Inc.}
\newblock Faqs about trends on twitter.
\newblock Accessed online at https://support.twitter.com/articles/101125, July
  2016.

\bibitem{SEC2015}
{U.S. Securities and Exchange Commission}.
\newblock Updated investor alert: Social media and investing --- stock rumors.
\newblock Accessed online at
  http://www.sec.gov/oiea/investor-alerts-bulletins/ia\_rumors.html, November
  2015.

\bibitem{varol2017online}
Onur Varol, Emilio Ferrara, Clayton~A Davis, Filippo Menczer, and Alessandro
  Flammini.
\newblock Online human-bot interactions: Detection, estimation, and
  characterization.
\newblock In {\em Proc. International Conference on Web and Social Media}.
  AAAI, 2017.
\newblock arXiv preprint arXiv:1703.03107.

\bibitem{varol2014evolution}
Onur Varol, Emilio Ferrara, Christine~L Ogan, Filippo Menczer, and Alessandro
  Flammini.
\newblock Evolution of online user behavior during a social upheaval.
\newblock In {\em Proceedings of the 2014 ACM conference on Web science}, pages
  81--90. ACM, 2014.

\bibitem{wang2015burst}
Senzhang Wang, Zhao Yan, Xia Hu, Philip~S Yu, and Zhoujun Li.
\newblock Burst time prediction in cascades.
\newblock In {\em Proceedings of the Twenty-Ninth AAAI Conference on Artificial
  Intelligence}, pages 325--331. AAAI Press, 2015.

\bibitem{warriner2013norms}
Amy~Beth Warriner, Victor Kuperman, and Marc Brysbaert.
\newblock Norms of valence, arousal, and dominance for 13,915 english lemmas.
\newblock {\em Behavior research methods}, pages 1--17, 2013.

\bibitem{weng2012competition}
L~Weng, A~Flammini, A~Vespignani, and F~Menczer.
\newblock Competition among memes in a world with limited attention.
\newblock {\em Scientific Reports}, 2, 2012.

\bibitem{weng2012virality}
L~Weng, F~Menczer, and YY~Ahn.
\newblock Virality prediction and community structure in social networks.
\newblock {\em Scientific reports}, 3:2522--2522, 2012.

\bibitem{weng2013virality}
Lilian Weng, Filippo Menczer, and Yong-Yeol Ahn.
\newblock Virality prediction and community structure in social networks.
\newblock {\em Scientific reports}, 3, 2013.

\bibitem{weng2013role}
Lilian Weng, Jacob Ratkiewicz, Nicola Perra, Bruno Gon{\c{c}}alves, Carlos
  Castillo, Francesco Bonchi, Rossano Schifanella, Filippo Menczer, and
  Alessandro Flammini.
\newblock The role of information diffusion in the evolution of social
  networks.
\newblock In {\em Proceedings of the 19th ACM SIGKDD international conference
  on Knowledge discovery and data mining}, pages 356--364. ACM, 2013.

\bibitem{wilson2005recognizing}
Theresa Wilson, Janyce Wiebe, and Paul Hoffmann.
\newblock Recognizing contextual polarity in phrase-level sentiment analysis.
\newblock In {\em Proceedings of the conference on Human Language Technology
  and Empirical Methods in Natural Language Processing}, pages 347--354. ACL,
  2005.

\bibitem{wu2007novelty}
Fang Wu and Bernardo~A Huberman.
\newblock Novelty and collective attention.
\newblock {\em Proceedings of the National Academy of Sciences},
  104:17599--17601, 2007.

\bibitem{yang2011patterns}
Jaewon Yang and Jure Leskovec.
\newblock Patterns of temporal variation in online media.
\newblock In {\em Proceedings of the fourth ACM international conference on Web
  search and data mining}, pages 177--186. ACM, 2011.

\bibitem{zhao2015enquiring}
Zhe Zhao, Paul Resnick, and Qiaozhu Mei.
\newblock Enquiring minds: Early detection of rumors in social media from
  enquiry posts.
\newblock In {\em Proceedings of the 24th International Conference on World
  Wide Web}, pages 1395--1405. International World Wide Web Conferences
  Steering Committee, 2015.

\end{thebibliography}

\end{document}